\documentclass[12pt,aps,prb,preprint]{revtex4}   

\usepackage{amsmath}    
\usepackage{amsfonts}  
\usepackage{amssymb}
\usepackage{graphicx}   
\usepackage{mathrsfs}
\usepackage{color}
\newcommand{\be}{\begin{equation}}
\newcommand{\ee}{\end{equation}}
\newcommand{\bea}{\begin{eqnarray}}
\newcommand{\eea}{\end{eqnarray}}
\newcommand{\nn}{\nonumber}
\newcommand{\Dd}{{\cal D}}

\begin{document}

\title
{On How the Scalar Propagator Transforms Covariantly in Spinless Quantum Electrodynamics}

\author{Yajaira Concha S\'anchez}
\affiliation{Facultad de Ingenier\'{\i}a Civil, Universidad
Michoacana de San Nicol\'as de Hidalgo, Av. Francisco J. M\'ujica
s/n, Ciudad Universitaria, Morelia, Michoac\'an, 58030, M\'exico.}

\author{Laura Xiomara Guti\'errez Guerrero}
\affiliation{Instituto de F\'{\i}sica y Matem\'aticas, Universidad
Michoacana de San Nicol\'as de Hidalgo, Edificio C-3, Ciudad
Universitaria, Morelia, Michoac\'an 58040, M\'exico.}

\author{Alfredo Raya}
\affiliation{Instituto de F\'{\i}sica y Matem\'aticas, Universidad Michoacana
de San Nicol\'as de Hidalgo, Edificio C-3, Ciudad Universitaria, Morelia, Michoac\'an
58040, M\'exico.}
\affiliation{Facultad de F\'{\i}sica, Pontificia Universidad Cat\'olica de Chile, Casilla 306, Santiago 22, Chile.}


\author{Victor M. Villanueva-Sandoval}
\affiliation{Instituto de F\'{\i}sica y Matem\'aticas, Universidad Michoacana
de San Nicol\'as de Hidalgo, Edificio C-3, Ciudad Universitaria, Morelia, Michoac\'an
58040, M\'exico.}

\begin{abstract}

Gauge covariance properties of the scalar propagator in
spinless/scalar quantum electrodynamics (SQED) are explored in the
light of the corresponding Landau-Khalatnikov-Fradkin
transformation (LKFT). These transformations are non perturbative
in nature and describe how each Green function of the gauge theory
changes under a variation of the gauge parameter. With a simple
strategy, considering the scalar propagator at the tree level in
Landau gauge, we derive a non perturbative expression for this
propagator in an arbitrary covariant gauge and three as well as
four space--time dimensions. Some relevant kinematical limits are
discussed. Particularly, we compare our findings in the weak
coupling regime with the direct one-loop calculation of the said
propagator and observe perfect agreement up to an expected gauge
independent term. We further notice that some of the coefficients
of the all-order expansion for the propagator are fixed directly
from the LKFT, a fact that makes this set of transformations
appealing over ordinary perturbative calculations in gauge
theories.

\end{abstract}

\maketitle

\section{Introduction}
Gauge principle has become the foundation of our modern
understanding of fundamental interactions among the basic
constituents of the Universe. Demanding the invariance of the
Lagrangian under local gauge transformations dictates the form of
the interactions among quarks and leptons in the celebrated
Standard Model of Particle Physics. The only abelian theory based
upon the gauge principle is the theory of electromagnetic
interactions --a historical account of the origins of gauge
invariance can be found in excellent reviews~\cite{roots}--. The
topic is so important that it is mandatory in many textbooks at
the graduate and advanced undergraduate levels. Classically,
Maxwell's equations which describe electromagnetic interactions
are usually solved for the auxiliary scalar and vector potentials
$\phi(\mathbf{x},t)$ and $\mathbf{A}(\mathbf{x},t)$ rather than
for the electric $\mathbf{E}(\mathbf{x},t)$ and magnetic field
$\mathbf{B}(\mathbf{x},t)$ themselves. These fields can later be
obtained from the potentials as follows~:
\begin{subequations}
\begin{eqnarray}
\mathbf{B}(\mathbf{x},t) &=& \nabla \times \mathbf{A}(\mathbf{x},t)\;,\\
\mathbf{E}(\mathbf{x},t) &=& -\nabla\phi(\mathbf{x},t) -\frac{\partial \mathbf{A}(\mathbf{x},t)}{\partial t}\;,
\end{eqnarray}
\end{subequations}
in cgs units. The choice of the electromagnetic potentials is not
unique, though the fields $\mathbf{E}(\mathbf{x},t)$ and
$\mathbf{B}(\mathbf{x},t)$ remain unaltered for every possible
equivalent configuration of the auxiliary
potentials~\cite{Jackson}. At the quantum level, gauge symmetry is
often introduced as a set of transformation rules for the
elementary fields --which represent particles-- of a given theory
that generate interaction terms and leave the resulting Lagrangian
invariant~\cite{Holstein,Noether}. Gauge invariance of a field
theory is of cardinal importance because it is intimately
connected to its renormalizability. Physical observables like the
components of the field strength
tensor~\cite{LagrangianGaugeTrans} or the $S$-matrix
elements~\cite{OpGauge} remain unchanged under gauge
transformations. Gauge symmetry is also manifest at the level of
field equations. In quantum electrodynamics (QED), for instance,
it is by virtue of gauge symmetry that different $(n+1)-$point
Green functions are related to $n-$point ones through the
so-called Ward-Green-Takahashi identities~\cite{WGTI}. Possibly
the best known of these identities relate the 3-point vertex
$\Gamma^\mu(k,p)$ to the 2-point (inverse) fermion propagator
$S_F^{-1}(p)$ in the form \be (k-p)_\mu
\Gamma^\mu(k,p)=S_F^{-1}(k)-S_F^{-1}(p)\;. \ee An enlarged set of
these identities, referred to as the Nielsen
identities~\cite{Nielsen}, have been used to demonstrate that the
poles of the propagators on the real time-like axis, corresponding
to the physical masses of the particles represented by these
fields, are gauge invariant quantities~\cite{Steele}. Furthermore,
to prove the renormalizability of a theory (namely, its predictive
power), both sets of identities should be valid order by order in
perturbation theory. At high energies, like those in which
collisions take place in the Large Hadron Collider (LHC) at CERN,
the theoretical predictions of Standard Model can be compared with
the experiment in a perturbative scheme that allows to formulate a
framework to systematically improve the predictions for scattering
cross sections, decay widths and other quantities.

There exist a different set of rules which specify the way Green
functions change under a gauge transformation. These rules carry
the name of Landau-Khalatnikov-Fradkin  transformations
(LKFTs)~\cite{LKF} and were first introduced in QED. Later on,
LKFT's have been re-discovered in the past decades by several
authors~\cite{LKFnew}. In deriving them, no assumption is  made on
the strength of the interaction, and therefore the transformations
are valid both in the strong and weak coupling regimes. This
feature is very important for theories of strong-interactions like
quantum chromodynamics (QCD) at low energies, that is, in the
realm of hadron physics. Understanding the nature of hadron
spectra is one of the most challenging problems in particle
physics simply because QCD becomes highly non-linear at low
energies. Therefore, very specialized techniques, which include
lattice simulations~\cite{lattice}, effective
theories~\cite{effective} and other approaches like the solution
of the field equations of a quantum field theory~\cite{esd} have
to be developed in order to gain and understanding  of confinement
and dynamical chiral symmetry breaking, two emergent phenomena of
QCD responsible of the nature of the hadron spectra. For detailed
reviews and recent applications of these equations, one can
consult the references~\cite{esd1}.

LKFTs have been employed in this context to enhance validity of
the predictions arising from the equations of motion of QED is
several space-time dimensions~\cite{LKFinSDE}. On the other hand,
these transformations also provide important information about the
all order perturbative expansion of the Green functions when some
perturbative form of these functions is used as input. In this
work we shall exploit the remarkable features of LKFTs to have a
glimpse into the non-perturbative structure of the scalar
propagator in SQED in 3 and 4 space--time dimensions whose
perturbative structure at the one loop level has been studied in
detail recently~\cite{Yajaira}.
We shall observe
that when the input we use in the LKFT approach is the
perturbative propagator at a given order $n_0$ in perturbation
theory in Landau gauge, all the gauge dependent terms at the
$(n_0+1)$-order get fixed by the weak coupling expansion of its
corresponding LKFT.
The paper is organized as follows: In the next section we
introduce the LKFT for the scalar propagator. We discuss the three
dimensional case in various interesting limits in Sect.~\ref{3D}.
Section~\ref{4D} is devoted to the 4D case, discussed in detail in
relevant kinematic limits. For the sake of comparison, we perform
an ordinary one-loop calculation of the scalar propagator and
contrast the prediction with those coming from the LKFT approach
in Sect.~\ref{1loop}. Our final remarks are presented in
Sec.~\ref{conclu}, and some calculations are discussed in an
appendix. At different stages of the discussion, crucial steps
that lead to our results are proposed as exercises for the
intrepid reader.

\section{Gauge Covariance and the Scalar Propagator }

Let us begin by recalling the Lagrangian of SQED. In order to keep
the notation as clear as possible, we do not consider the
space-time dependence of the fields in the following. Free charged
scalar particles (with the fundamental charge $e$) of mass $m$,
represented by the fields $\phi$ and $\phi^*$ are described by the
Lagrangian
 \be {\cal L}_{\rm scalar}= \frac{1}{2}
 \left[(\partial_\mu\phi)(\partial^\mu\phi)^* \right]
 - \frac{1}{2} \, m^2 \, \phi^* \phi\;.
 \ee
Next, we allow these fields to couple electromagnetically with
photons. This is achieved by promoting the ordinary derivatives to
the covariant ones
 \be
 \partial_\mu \to \mathscr{D}_\mu=\partial_\mu+ieA_\mu\;,
 \ee
where the vector field $A_\mu$ represents the electromagnetic
potential, which is associated with the photon wavefunction in the
quantum theory. We further add the electromagnetic kinetic term
(Maxwell term), \be {\cal L}_\gamma =
-\frac{1}{4}F_{\mu\nu}F^{\mu\nu}\;, \ee with
$F^{\mu\nu}=\partial^\mu A^\nu - \partial^\nu A^\mu$ representing
the field strength tensor, as usual. Finally, for the consistency
of the second quantized theory, we must add two terms. The first
additional one is the covariant gauge fixing term, which gets rid
of spurious degrees of freedom in a covariant manner, and the
second is the scalar self-interaction quartic term, which is
required for the renormalizability of the theory.  Therefore, we
shall be working with the Lagrangian:
 \be {\cal L}= \frac{1}{2}
 \left[(\mathscr{D}_\mu\phi)(\mathscr{D}^\mu\phi)^* \right]
 -\frac{1}{2} \, m^2 \, \phi^* \phi -\frac{1}{4}F_{\mu\nu}F^{\mu\nu}-
 \frac{1}{2\xi}(\partial_\mu A^\mu)^2 -\frac{\lambda}{4}\left(
 \phi^*\phi\right)^2, \label{lag}
 \ee
where  $\lambda$ the coupling for the scalar self-interaction and
$\xi$ is the covariant gauge parameter. The above
Lagrangian~(\ref{lag}) describes, for instance, the
electromagnetic and self-interactions of charged scalars like
pions at low energies where they can be considered as point
particles. Observe that the electromagnetic dynamics between
scalar particles is oblivious of the self interaction part other
than the technicalities on renormalization. As we are focussing
only on the gauge covariance relations, the non-gauge 4-point self
interaction is irrelevant to our purpose. Hence, in the remainder
of the article, we set $\lambda=0$.

The next step in our program is to identify the most important
quantities that are relevant for our discussion, namely, the
propagators. The free photon propagator associated to the
Lagrangian~(\ref{lag}) is \be
\Delta_{\mu\nu}^{(0)}(q;\xi)=-\frac{1}{q^2}\left(g_{\mu\nu}-\frac{q_\mu
q_\nu}{q^2} \right) +\xi \frac{q_\mu
q_\nu}{q^4}\;.\label{photonprop} \ee We explicitly label the
propagators with the covariant gauge parameter $\xi$ (a real
number) as we would be interested in their expression in different
gauges. The photon propagator~(\ref{photonprop}) changes from
gauge to gauge simply by choosing different values of  $\xi$. Some
useful gauges are the Feynman gauge ($\xi=1$), amenable for
perturbative calculations; the Landau gauge ($\xi=0$), which
retains the transverse nature of the photons in a manifestly
covariant form;  or Yennie gauge ($\xi=3$), where the mass shell
renormalization scheme can be implemented without introducing
spurious infrared divergences. The tree level scalar propagator,
in a $d$-dimensional Minkowski space-time, has the form
 \be
 iD^{(0)}_d(p;\xi)=  \frac{i}{p^2-m^2},\label{tree}
 \ee where the
Feynman prescription $+i\varepsilon$ is understood, but omitted.
As stated by the gauge principle, a change in the photon
propagator~(\ref{photonprop}) should be compensated by a change in
the scalar propagator~(\ref{tree}) to retain the gauge invariance
of the Lagrangian. There exist a set of transformation rules that
specify the explicit form in which all the Green functions in our
theory change under local gauge transformations. These are called
Landau-Khalatnikov-Fradkin transformations (LKFT)~\cite{LKF} and
have been discovered and re-discovered during the past 60
years~\cite{LKFnew}. These transformations have the simplest
structure in Euclidean coordinate space. Therefore, we relate the
propagator in $d$-dimensional Euclidean coordinate and momentum
spaces, which we represent by the symbols $\Dd_d(x;\xi)$ and
$\Dd_d(p;\xi)$, respectively, through
\begin{subequations}
\bea
\Dd_d(x;\xi)&=& \int \frac{d^dp}{(2\pi)^d} \ {\rm e}^{-ip\cdot x}\Dd_d(p;\xi)\;,\label{f2x}\\
\Dd_d(p;\xi)&=& \int d^dx \ {\rm e}^{ip\cdot x}\Dd_d(x;\xi)\;, \label{f2p}
\eea
\end{subequations}
The LKFT relating the coordinate space scalar propagator in Landau
gauge to the one in an arbitrary covariant gauge in arbitrary
space-time dimensions $d$ reads~: \bea \Dd_d(x;\xi) =
\Dd_d(x;0){\rm e}^{-i [\Delta_d(0)-\Delta_d(x)]} \;,\label{LKprop}
\eea where \bea
\Delta_d (x)&=&-i \xi e^2 \mu^{4-d}\int \frac{d^dp}{(2\pi)^d} \frac{{\rm e}^{-ip\cdot x}}{p^4}\nn\\
&=&-\frac{i \xi e^2}{16 (\pi)^{d/2}} (\mu
x)^{4-d}\Gamma\left(\frac{d}{2}-2\right) \;.\label{deltad} \eea
Here, $\mu$ is a mass scale introduced for dimensional purposes;
it ensures that in every dimension $d$, the coupling $e$ is
dimensionless. $\Gamma(x)$ is the gamma function.
\begin{quote}
{\bf Exercise:} In order to verify the result in Eq.~(\ref{deltad}),
use hyperspherical coordinates such that $d^dp=p^{d-1}dp \
\sin^{d-2}\theta d\theta \ \Omega_{d-2}$ where
$\Omega_{d-2}=2\pi^{(d-1)/2}\Gamma\left((d-1)/2\right)$ is the
solid angle in $d-2$ dimensions. Orient your reference frame such
that $p\cdot x =p x \cos\theta$.
\end{quote}

With these definitions, we are ready to study the gauge covariance
of the scalar propagator from its LKFT. The strategy is as
follows~:
 \begin{enumerate}
 \item[(i)] Start from the bare propagator in momentum space and
 Fourier transform it to coordinate space;
 \item[(ii)] Apply the LKFT;
 \item[(iii)]  Fourier transform it back to momentum space.
 \end{enumerate}
We proceed to perform this exercise below and encourage the reader to carry
out all the steps to master the strategy.

\section{Three-dimensional case}~\label{3D}

Three-dimensional theories offer more than a mere simplification
for academic ``training''. These emerge naturally as the limit of
the corresponding four-dimensional theories at very high
temperatures~\cite{pisarski} and/or densities, in the presence of
ultra strong magnetic fields~\cite{landau} and other {\em extreme}
conditions. Moreover, these theories are appealing by themselves
because their internal richness associated, for instance, to
fractional spin-statistics (anyons)~\cite{anyon} and generalized
parity properties~\cite{parity}. In this Section, we are
interested in the gauge covariance properties of the scalar
propagator in a three-dimensional Euclidean space.

Setting $d=3$ and identifying $e\mu \to e$,  we can readily
observe from Eq.~(\ref{deltad}) that \be
\Delta_3(0)-\Delta_3(x)=-iax\;, \ee where $a=\alpha\xi/2$ and
$\alpha=e^2/(4\pi)$ is the usual electromagnetic coupling, which
has dimensions of mass. This implies that the LKFT~(\ref{LKprop})
is given by \bea \Dd_3(x;\xi)={\rm
e}^{-ax}\Dd_3(x;0).\label{LKFT3} \eea

Following our strategy,  we consider the scalar propagator in
$d=3$ in Euclidean space, i.e., our input is
 \be
 \Dd_3(p;0)=-\frac{1}{p^2+m^2}\;,\label{bareS}
 \ee
This expression is the same in every gauge. However, for
calculational convenience, we consider it in the Landau gauge.
Thus we can perform the Fourier transform~(\ref{f2x}) using
spherical coordinates. Writing $d^3p=p^2dp \sin\theta d\theta
d\varphi$ and orienting the reference frame such that $p\cdot x =
p x \cos\theta$, then we have that \bea
\Dd_3(x;0)&=&-\frac{1}{(2\pi)^3} \int_0^\infty \frac{dp\ p^2}{p^2+m^2}\int_0^\pi d\theta \sin\theta \ {\rm e}^{-ipx\cos\theta}\int_0^{2\pi} d\varphi \nn\\
&=&-\frac{2}{(2\pi)^2} \int_0^\infty \frac{dp\ p^2}{p^2+m^2} \frac{\sin(px)}{px}\nn\\
&=& -\frac{1}{4\pi x}{\rm e}^{-mx}\;. \eea The LKFT is straight
forward to perform. It would simply shift the argument of the
exponential in the above expression by the amount $-ax$. Then we
are only left with the inverse Fourier transform, which readily
leads to \be \Dd_3(p;\xi)=-\frac{1}{p^2+\left(m+a\right)^2}\;.
\label{LKFresults3D} \ee The expression (\ref{LKFresults3D})
yields the non-perturbative form of the scalar propagator in an
arbitrary covariant gauge. {\em An important advantage of the LKFT
over ordinary perturbative calculation is that the weak coupling
expansion of this transformation already fixes some of the
coefficients in the all order perturbative expansion of the
fermion propagator (see, for
example,~\cite{LKFper,LKFperb,LKFperc,LKFper2}).} Let us consider
some special cases of the expression~(\ref{LKFresults3D}).

\subsection{Massless Case}

At very high energies, the mass of the particles can be neglected,
and therefore, the propagator has a simple form. In the massless
case, the LKF result~(\ref{LKFresults3D}) reduces to \be
\Dd_{3,{\rm massless}}(p;\xi)=-\frac{1}{p^2+a^2}\;.
\label{LKF3Dmassless} \ee Apparently, the LKFT has generated
non-perturbatively an explicitly gauge dependent mass $m_{\rm
LKF}=a$ for massless scalar particles. This is a feature of LKFT.
However this cannot be correct on physical grounds: The poles of
propagators must be gauge invariant quantities~\cite{Steele}. The
reason for this apparent mishap can be traced back to our
incomplete input: We are estimating the full scalar propagator
only in terms of its tree-level counterpart. Had we started with
the full scalar propagator, the poles we would obtain are of
course gauge independent. In fact, the more we refine our starting
guess with a multi-loop scalar propagator, the spurious gauge
dependent pole is washed away systematically~\cite{LKFper}.

\subsection{Weak coupling expansion}

From Eq.~(\ref{LKFresults3D}), the inverse propagator is explicitly
\be
\Dd^{-1}_3(p;\xi)=-\left[p^2+m^2 + m\alpha\xi+\frac{\alpha^2\xi^2}{4} \right].\label{LK3dweak}
\ee
This expression should be contrasted against the one-loop perturbative result of Sect.~\ref{1loop}. However, we can establish some general conclusions about
the predictive power of LKFT.  For this purpose, let us perform a
weak coupling expansion of the Fourier transform of the
propagator~(\ref{LKprop}) in momentum space,
 \be \Dd_d(p;\xi)=
 a_0\alpha^0+ (a_1\xi+\mathfrak{b}_1)\alpha
 +(a_2\xi^2+\mathfrak{b}_2\xi+{\rm {\bf c}}_2)\alpha^2+\ldots \;.
 \ee
We notice that the LKFT fixes the coefficients of the form
$(\alpha\xi)^n$. In our three-dimensional example, knowledge of
$a_0$ allows the LKFT to fix \be a_0=-(p^2+m^2)\;, \quad
a_1=-m\;,\quad a_2=-\frac{1}{4}\;, \quad a_n=0 \ \forall \ n\ge
3.\label{perLK} \ee However, starting from the tree-level
propagator, we cannot infer the coefficients of the crossed-terms
such as $\mathfrak{b}_1$, $\mathfrak{b}_2$, ${\rm {\bf c}}_2$ and
so on, a fact that holds true in arbitrary space-time dimensions
and also for the spinor case, as pointed out in~\cite{LKFper2}.
Even more, when our starting input is a ${\cal O}(\alpha^n)$
perturbative propagator, all  the terms of the form
$\alpha^{n+i}\xi^i$ would already get fixed, as well as those with
higher powers of $\xi$ at a given order in $\alpha$ after the weak
coupling expansion of the results obtained by applying the
corresponding LKFT~\cite{LKFperc}.

\subsection{Momentum-space representation}

Most calculations of cross-sections or decay rates in particle
physics are best carried out in momentum space. A reason for that
is that the energy-momentum and other conservation laws can be
expressed more transparently in momentum as opposed to coordinate
spaces. Moreover, there are situations in which the Fourier
transformation involved in the LKFT strategy cannot be expressed
in a  closed form.  However, in $d=3$ we can have en explicit
momentum-space representation for the LKFT (see the last article
in~\cite{LKFinSDE}). Let us assume that by some means we are given
a general form of the scalar propagator in momentum space in
Landau gauge. That is, let us assume that all we know is some
multi-loop or non perturbative form of $\Dd_3^\star(p;0)$ (it
might well be numerical or a very complicated function of its
arguments). Then, from Eq.~(\ref{f2x}), using spherical
coordinates, we have that \bea
\Dd_3^\star(x;0) &=& \frac{1}{(2\pi)^3} \int_0^\infty dp\ p^2 \Dd_3^\star(p;0) \int_0^\pi d\theta\ \sin\theta\ {\rm e}^{-ipx\cos\theta}\nn\\
&=& \frac{2}{(2\pi)^2} \int_0^\infty dp\ p^2 \Dd_3^\star(p;0) \frac{\sin(p x)}{p x}\nonumber\\
&=& \frac{1}{2\pi^2x}\int_0^\infty dp\ p \sin(p x)
\Dd_3^\star(p;0)\;, \eea such that \be \Dd_3^\star(x;\xi)=
\frac{{\rm e}^{-a x}}{2\pi^2x}\int_0^\infty dp\ p \sin(p x)
\Dd_3^\star(p;0)\;.\label{LKFcoord} \ee On the other hand, from
Eq.~(\ref{f2p}) and considering a general expression for
$\Dd_3^\star(x;\xi)$, a parallel reasoning yields \be
\Dd_3^\star(k;\xi)= 4\pi \int_0^\infty dx\ x^2
\Dd_3^\star(x;\xi)\frac{\sin(k x)}{k x}\;. \ee Inserting
$\Dd_3^\star(x;\xi)$ from Eq.~(\ref{LKFcoord}), we have that \be
\Dd_3^\star(k;\xi)= \frac{2}{\pi} \int_0^\infty dx \int_0^\infty
dp\ \frac{p}{k} \sin(k x) \sin(p x) {\rm e}^{-a x}
\Dd_3^\star(p;0)\;. \ee The integral is convergent for $a>0$, and
thus, exchanging the order of integration, we arrive at the
momentum space representation of the LKFT for the scalar
propagator, \be \Dd_3^\star(k;\xi)= \frac{a}{\pi k} \int_0^\infty
dp\ p \Dd_3^\star(p;0) \left[\frac{1}{a_-}-\frac{1}{a_+} \right]
\;, \label{momentum} \ee with \be a_\pm = a^2+(k\pm p)^2\;. \ee
This expression allows us to avoid the use of the coordinate-space
representation of a multi-loop or non-perturbative scalar
propagator.
\begin{quote}
{\bf Exercise:} Show that inserting the propagator~(\ref{bareS}) into the formula~(\ref{momentum}), we obtain the same result~(\ref{LKFresults3D}).
\end{quote}

This completes our discussion on the 3-dimensional case. Below we
consider the 4-dimensional case.

\section{Four-dimensional case}~\label{4D}

Let us now consider the ordinary theory in $d=4$. Notice that the
difference  $\Delta_4(0)-\Delta_4(x)=0$, thus suggesting that the
transformation becomes trivial. However, the scalar propagator
itself is divergent in this case, and therefore, the propagator
and its LKFT should be regulated. A favorite procedure  is to
regularize with the dimension by considering $d=4-2\epsilon\equiv
4^*$ space--time dimensions in the limit $\epsilon\to 0\
$~\cite{DimReg}. Then, naturally appears a cut-off limit $x_{\rm
min}$ which in turn indicates that \be \Delta_{4^*}(x_{\rm
min})-\Delta_{4^*}(x)=-i\ln\left|\frac{x^2}{x_{\rm
min}^2}\right|^\nu\;,\label{LKF4} \ee where $\nu = a/(2\pi)$ and
$x_{\rm min}$ is an IR cut-off in coordinate space introduced to
regulate the transformation.
\begin{quote}
{\bf Exercise:} Show that the definition~(\ref{deltad}) for $d=4^*$ leads to~(\ref{LKF4}).
\end{quote}
In this way,  the LKFT is
\bea
\Dd_{4^*}(x;\xi)=\Dd_{4^*}(x;0)\left(\frac{x^2}{x^2_{\rm min}} \right)^{-\nu}.\label{LKFT4}
\eea

At this point, we have all the ingredients to follow the LKFT
strategy again. Let us begin from the expression~(\ref{tree}) for
the tree level scalar propagator, but now in a four-dimensional
Euclidean space. It is safe to take directly $d=4$, since the
Fourier transform integrals are convergent. Again, we use
hyperspherical coordinates and write $d^4p = p^3dp\sin^2\theta
d\theta \sin\psi d\psi d\varphi$. Orienting the reference frame
such that $p\cdot x=px\cos\theta$, the tree-level propagator in
coordinate space is
 \bea
 \Dd_4(x;0)&=& -\frac{1}{(2\pi)^4}\int_0^\infty \frac{dp \ p^3}{p^2+m^2}\int_0^\pi d\psi\sin^2\theta \ {\rm e}^{-ipx\cos\theta} \int_0^\pi d\psi \sin\psi \int_0^{2\pi}d\varphi\nonumber\\
&=&- \frac{4\pi}{(2\pi)^4}\int_0^\infty \frac{dp \ p^3}{p^2+m^2}\int_0^\pi d\theta\sin^2\theta\  {\rm e}^{-ipx\cos\theta}\nn\\
&=& -\frac{\pi}{4\pi^3}\int_0^\infty \frac{dp \ p^3}{p^2+m^2}\frac{J_1(px)}{px}\nn\\
&=& -\frac{m}{4\pi^2x} K_1(mx)\;, \eea where $J_1(z)$ and $K_1(z)$
are, respectively, the Bessel function of the first kind and order
one and the modified Bessel function of the second kind and order
one. In a general covariant gauge, the propagator acquires the
form \be \Dd_4(x;\xi)= -\frac{m}{4\pi^2x}
K_1(mx)\left(\frac{x^2}{x^2_{\rm min}}
\right)^{-\nu}\;,\label{4Dx} \ee and thus the non-perturbative
propagator derived through LKFT in momentum space is \be
\Dd_4(p;\xi)=-\frac{1}{m^2}\left(
\frac{m^2}{\Lambda^2}\right)^{\nu} \Gamma(1-\nu) \Gamma(2-\nu)\
_2F_1\left(1-\nu,2-\nu;2;-\frac{p^2}{m^2}\right)\;,\label{LKFresult4D}
\ee where $\Lambda^2 =4/x_{\rm min}^2 $ and $_2F_1(a,b;c;x)$ is
the hypergeometric function.
\begin{quote}
{\bf Exercise:} Show that the 4D Fourier transform~(\ref{f2p}) of Eq.~(\ref{4Dx}) leads to Eq.~(\ref{LKFresult4D}).
\end{quote}
Let us consider some particular limits of this expression.

\subsection{Massless Case}

 Let us study the kinematical regime $m=0$ from LKFT. Notice that the argument of the hypergeometric function is divergent, and therefore the limit cannot be taken directly. However, using the property
\be _2F_1(a,b;c;z)= (1-z)^{-a}\
_2F_1\left(a,c-b;c;\frac{z}{z-1}\right)\;, \ee we can write the
propagator~(\ref{LKFresult4D}) in the equivalent form \be
\Dd_4(p;\xi)=-\frac{1}{\Lambda^{2\nu}}\Gamma(1-\nu) \Gamma(2-\nu)
(p^2+m^2)^{-1+\nu} \
_2F_1\left(1-\nu,\nu;2;\frac{p^2}{p^2+m^2}\right)\;. \ee Now the
limit $m\to 0$ can be taken safely. The argument of the
hypergeometric function becomes 1, and recalling that \be
_2F_1(a,b;c;1)=\frac{\Gamma(c)\Gamma(c-a-b)}{\Gamma(c-a)\Gamma(c-b)}\;,
\ee we reach the final expression for the massless scalar
propagator \be \Dd_{\rm 4, massless}(p;\xi)=
-\frac{1}{p^2}\frac{\Gamma(1-\nu)}{\Gamma(1+\nu)}\left(\frac{p^2}{\Lambda^2}
\right)^\nu \;. \ee If we further consider the limit of weak
coupling ($\nu\to 0$), at the leading order we have that \be
\Dd_{\rm 4, massles}(p;\xi)\stackrel{\nu\to
0}{=}-\frac{1}{p^2}\Bigg[1+\nu\left(2\gamma_E
+\log\left(\frac{p^2}{\Lambda^2}\right)\right)  \Bigg]+{\cal
O}(\nu^2)\;,\label{masslessweakLKF} \ee where $\gamma_E$ is the
Euler constant. This expression will allow to establish a
comparison against ordinary perturbation theory in
Sect.~\ref{1loop}, but for the time being, let us explore some
other interesting limits.


\subsection{Static Limit}

The static limit is achieved when the scalar mass is much larger
as compared to the momentum flowing through it. Let $z=p^2/m^2$
and let us consider the case as $z\to 0$. We can make use of the
well known expansion of the hypergeometric function for small
argument \be _2F_1(a,b;c;x)= 1-\frac{ab}{c}x+{\cal O} \left(
x^2\right)\;.\label{hypstat} \ee Therefore, the propagator becomes
\bea
\Dd_{\rm 4, static}(p;\xi)&=& -\frac{1}{m^2}\left(1-z\right)\nn\\
&&+\frac{\nu}{m^2}\left[1+2\gamma_E-\ln\left|\frac{m^2}{\Lambda^2}\right|
+z\left(-\frac{5}{2}+2\gamma_E+ \ln\left|\frac{m^2}{\Lambda^2}\right|\right) \right]\;.
\eea
For the sake of comparison, it is better to consider the inverse propagator instead. In the present case, we have that
\be
\Dd_{\rm 4, static}^{-1}(p;\xi)= -\frac{1}{m^2}\left(1-z\right)+\frac{\nu}{m^2}\left[1-\frac{z}{2}-(1+z)\left(
2\gamma_E+\ln\left|\frac{m^2}{\Lambda^2}\right|\right) \right]\;.
\ee
Next we consider the weak coupling regime with arbitrary mass scale.

\subsection{Weak coupling expansion}
In the weak coupling regime, we require to expand the hypergeometric functions in Eq.~(\ref{LKFresult4D}) in terms of its parameters. The task is complicated and out of the scope of this article. Besides, there exist
analytical~\cite{HypExp} and numerical~\cite{NumExp} tools to implement such an expansion in symbolic manipulation systems. Therefore we content ourselves by quoting the expansion, when $\varepsilon\to 0$, of~\cite{LKFperb}
\begin{equation}
_2F_1(1-\varepsilon,2-\varepsilon,2;x)\simeq\frac{1}{1-x}\Bigg[1+\varepsilon\Bigg\{1+\frac{1+x}{x}\ln(1-x) \Bigg\} \Bigg]\;,
\end{equation}
such that
\bea
\Dd(p;\xi)_{\rm 4, weak}&=& -(p^2+m^2)\nn\\
&+&(p^2+m^2)\nu\left[\ln\left|\frac{m^2}{\Lambda^2} \right| +2\gamma_E+\left(1-\frac{m^2}{p^2} \right)\ln\left|1+\frac{p^2}{m^2} \right|\right]\;.\label{1loopLKF}
\eea

We next perform a perturbative calculation of the scalar
propagator and compare against the findings of this and the
previous Sections in order to understand the working of LKFT.

\section{One-loop perturbative scalar propagator}~\label{1loop}

\begin{figure}[t]
\includegraphics[angle=-90]{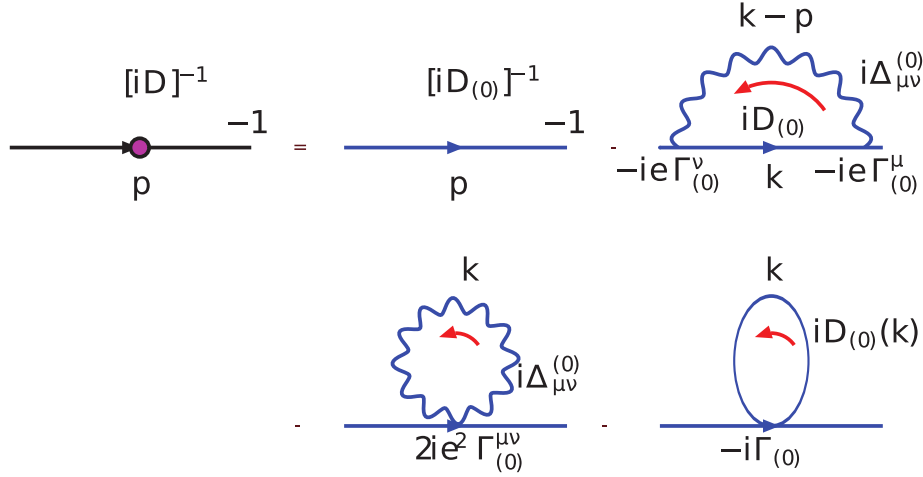}
\caption{One-loop correction to the scalar propagator.}
\label{fig1}
\end{figure}

In this section we carry out a one-loop calculation of the scalar
propagator with the standard approach. From the
Lagrangian~(\ref{lag}), we can read the corresponding Feynman
rules
\begin{itemize}
\item For every vertex of one photon and two scalars $\phi_A$ and $\phi_B^*$, with momentum $P_A$ and $P_B$, respectively, we consider the factor $-ie\Gamma_0^\mu=-ie(P_A+P_B)^\mu$.
\item For the vertex of 2 photons labeled by the Lorentz indices $\mu$ and $\nu$, and two scalars $\phi_A$ and $\phi_B^*$, with momentum $P_A$ and $P_B$, respectively, we add a factor $2ie^2\Gamma_0^{\mu\nu}=2ie^2g^{\mu\nu}$.
\item For a four-scalar vertex, a factor $-i\Gamma_0=-i\lambda$.
\end{itemize}
Moreover, for the internal lines we use the photon
propagator~(\ref{photonprop}) and the scalar
propagator~(\ref{tree}) in the diagrams shown in Fig.~\ref{fig1},
corresponding to the one-loop correction to scalar propagator.
In arbitrary space-time dimensions $d$, it corresponds to the following expression in Minkowski space
\bea
[iD^{(1)}_d(p;\xi)]^{-1} &=& [iD^{(0)}_d(p;\xi)]^{-1} - \int\frac{d^dk}{(2\pi)^d} [2ie^2\Gamma_0^{\mu\nu}][i\Delta_{\mu\nu}^{(0)}(k;\xi)]
\nn\\
&&
 -\int\frac{d^dk}{(2\pi)^d} [-ie\Gamma_0^\mu][iD_d^{(0)}(k;\xi)][-ie\Gamma_0^\nu] [i\Delta_{\mu\nu}^{(0)}(k-p;\xi)]\nn\\
&&
 - \int\frac{d^dk}{(2\pi)^d} [iD_d^{(0)}(k;\xi)][-i\Gamma_0]\nn\\
&=&[iD^{(0)}_d(p;\xi)]^{-1}-I_1-I_2-I_3\;. \eea The integrals are
divergent and thus require a regularization. It is customary to
proceed within the dimensional regularization
scheme~\cite{DimReg}, and we adopt this view for two different
purposes: (i) To stick with the ordinary conventions of
perturbative calculations in quantum field theory and (ii) to
provide a link between the cut-off regularization of the LKFT
scheme with this procedure. We start noticing that $I_1=0$, so,
writing the explicit form of the remaining terms, we have that
\bea
[iD^{(1)}_d(p;\xi)]^{-1} &=& [iD^{(0)}_d(p;\xi)]^{-1}-i\lambda\int\frac{d^dk}{(2\pi)^d}\frac{1}{k^2-m^2}\nn\\
&&\hspace{-15mm}
 -\int\frac{d^dk}{(2\pi)^d} (k+p)^\nu \frac{1}{k^2-m^2} (k+p)^\mu \frac{1}{q^2} \left[g_{\mu\nu}+(\xi-1)\frac{q_\mu q_\nu}{q^2} \right]\;.\nn\\
 && \hspace{-15mm} = [iD^{(0)}_d(p;\xi)]^{-1}+(1-i\lambda) J_{01} +2(m^2+p^2)J_{11}+2J_{10}+(m^2-p^2)^2J_{21}\;,
\eea where the master integral is
 \bea
 J_{np}&=& \int d^dk \frac{1}{(q^2)^n (k^2-m^2)^p} \nn\\
 && \hspace{-5mm} = (-1)^{n+p} i \pi^\frac{d}{2} (m^2)^{\frac{d}{2}-n-p}
 \frac{\Gamma\left(n+p-\frac{d}{2}\right) \Gamma\left(\frac{d}{2}-n\right)}{\Gamma(p)\Gamma\left(\frac{d}{2} \right)}
\ _2F_1 \left(-\frac{d}{2}+n+p,n;\frac{d}{2}; \frac{p^2}{m^2}
\right).\label{master} \eea Details of the calculation of the
above master integral~(\ref{master}) are presented in the
appendix, because they correspond to standard perturbative
calculations in quantum field theory. Upon substituting the
appropriate values of $n$ and $p$, we are  led to the final result
for the one-loop correction to the scalar propagator, \bea
[iD^{(1)}_d(p;\xi)]^{-1} &=& p^2-m^2-\left( \frac{m^2}{4\pi}\right)^\frac{d}{2}\frac{1}{m^2}\Gamma \left( 1-\frac{d}{2}\right)\nn\\
&&\times
\Bigg\{ e^2 \Bigg[ 1-2 \frac{(m^2+p^2)}{m^2}\ _2F_1\left(2-\frac{d}{2}, 1; \frac{d}{2}; \frac{p^2}{m^2}\right)\nn\\
&& + (1-\xi) \frac{(m^2-p^2)}{m^4}\ _2F_1\left(3-\frac{d}{2}, 2;\frac{d}{2};\frac{p^2}{m^2} \right)\Bigg] +\lambda\Bigg\}.
\label{1loopd}
\eea
Let us consider some particular cases.

\subsection{Three-dimensional case}

Setting $d=3$ in Eq.~(\ref{1loopd}) reduces the hypergeometric
functions to transcendental, non divergent expressions, which
after some manipulations can be expressed as
 \bea
 [iD^{(1)}_3(p;\xi)]^{-1} = p^2-m^2
 + \alpha\left[m-\frac{2(m^2+p^2)}{m}\sqrt{\frac{m^2}{p^2}}\tanh^{-1}\sqrt{\frac{m^2}{p^2}}
 +(1-\xi)m\right]\;.
 \eea
In order to compare against the results arising from the LKFT, we
perform a Wick rotation to Euclidean space, such
that\footnote{Note that this form tells us that $\xi=2$ is even a
better starting gauge than the Landau gauge because the one-loop
effect is minimized in the former gauge, as noted by Delbourgo and
Keck~\cite{del1}.}.
 \bea
 [i\Dd^{(1)}_3(p;\xi)]^{-1} = -p^2-m^2
  +\alpha\left[m-2(m^2-p^2)I(p,m) +(1-\xi)m\right]\;,\label{1l3dmass}
 \eea
where we have defined \be
I(p,m)=\frac{1}{p}\arctan{\left(\frac{p}{m} \right)}\;.\label{Ip}
\ee The expression~(\ref{1l3dmass}) is in agreement with its
counterpart~(\ref{LK3dweak}) to the leading term. The gauge
independent terms cannot be derived from LKFT, as pointed our
before. Moreover, in the massless limit, the propagator becomes
\be \Dd^{(1)}_3(p;\xi) = \frac{1}{p^2+\alpha\pi p}\;,\label{1l3d}
\ee which softens its infrared divergence from $1/p^2$ to $1/p$.
We observe that the pole of this propagator is independent of
$\xi$.  Let us remark, as noticed earlier, that all the terms of
the form $(\alpha\xi)^n$ are in agreement with the LKFT
result~(\ref{LKF3Dmassless}).

\begin{quote}
{\bf Exercise:} We encourage the students to use the
expression~(\ref{1l3d}) as input in the formula~(\ref{momentum})
to find the non perturbative form of the propagator in different
gauges.
\end{quote}

This concludes the discusion of the perturbative three-dimensional case. Next we consider the four-dimensional case.

\subsection{Four-dimensional case}

In $d=4$, in the expression for the propagator~(\ref{1loopd}), the $\Gamma$ functions are divergent. We
can easily expand these functions in powers of $\epsilon$ by taking $d=4^*$.
Let us observe that when $d=4$, the hypergeometric functions reduce to
\begin{subequations}
\bea
_2F_1\left(2-\frac{d}{2},1;\frac{d}{2};\frac{p^2}{m^2} \right) &=& _2F_1\left(0,1;2;\frac{p^2}{m^2} \right) \ = \ 1\label{F1}\\
_2F_1\left(3-\frac{d}{2},2;\frac{d}{2};\frac{p^2}{m^2} \right) &=&
_2F_1\left(1,2;2;\frac{p^2}{m^2} \right) \ = \
\frac{m^2}{p^2+m^2}\;,\label{F2} \eea
\end{subequations}
and thus the only divergence comes from
\begin{subequations}
\bea
\Gamma\left(2-\frac{d}{2} \right) &=& \Gamma(\epsilon) \ = \ \frac{1}{\epsilon}+ \mbox{finite}\label{g1}\\
\Gamma\left(\frac{d}{2} -2\right) &=& \Gamma(-\epsilon) \ = \ -\frac{1}{\epsilon}+ \mbox{finite}\label{g2}\\
\Gamma\left(1-\frac{d}{2} \right) &=& \Gamma(-1+\epsilon) \ = \ -\frac{1}{\epsilon}+ \mbox{finite}\;, \label{g3}
\eea
\end{subequations}
such that when $d=4^*$, \be
\left[D_{4^*}^{(1)}(p;\xi)\right]^{-1}=
p^2-m^2-\frac{\alpha}{4\pi\epsilon}
\left[m^2+2p^2-(1-\xi)(m^2-p^2) \right] +
\mbox{finite.}\label{drlead} \ee This expression encodes the
leading divergence, which corresponds to the divergence of the
propagator when $\Lambda\to \infty$ in Eq.~(\ref{1loopLKF}). We
can improve on our understanding of the propagator by identifying
the finite piece in Eq.~(\ref{drlead}). For this purpose, we
resort to the software~\cite{HypExp,NumExp} to expand the
hypergeometric functions in $d=4^*$. Here we simply quote the
result and leave the details as an exercise to the reader, \bea
\left[\Dd_{4^*}^{(1)}(p;\xi)\right]^{-1}&=& p^2-m^2-\frac{m^2}{(4\pi^2)}\frac{1}{\epsilon} \Bigg[ 4\pi\alpha+\lambda+\frac{4\pi\alpha}{m^2} \bigg[ -2(m^2+p^2)+(1-\xi)(m^2-p^2)\bigg]\Bigg] \nn\\
&&+\frac{m^2}{(4\pi)^2} \Bigg[ C \Bigg( 4\pi\alpha+\lambda+\frac{4\pi\alpha}{m^2} \bigg[ -2(m^2+p^2)+(1-\xi)(m^2-p^2)\bigg] \Bigg) \nn\\
&&+ 4\pi\alpha \Bigg[ -2\frac{(m^2+p^2)}{m^2}\left(1-\frac{p^2-m^2}{p^2}\ln\left(1-\frac{p^2}{m^2} \right) \right) \Bigg] \nn\\
&&-4\pi\alpha (1-\xi)\frac{(m^2-p^2)}{m^2}\left(1-\frac{p^2-m^2}{p^2}\ln\left(1-\frac{p^2}{m^2} \right) \right) \Bigg\}\;,
\label{arbm}
\eea
where
\be
C=1-\gamma_E-\ln\left(\frac{m^2}{4\pi} \right)\;.
\ee
It is not difficult to convince oneself that the gauge dependent terms in Eq.~(\ref{arbm}) exactly match those in Eq.~(\ref{1loopLKF}), and that the only difference comes from the $\xi$-independent terms, a difference allowed by the structure of LKFTs.

\begin{quote}
{\bf Exercise:} We encourage graduate students to perform the intermediate steps to obtain the expression~(\ref{arbm}). The expansion of the hypergeometric functions is not trivial, so we provide the following results
\begin{subequations}
\label{hypergeo}
\bea
_2F_1\left(\epsilon,1; 2-\epsilon;\frac{p^2}{m^2}\right)&=& 1+\epsilon\Bigg[ 1-\frac{p^2-m^2}{p^2}\ln\left(1-\frac{p^2}{m^2} \right)\Bigg]+{\cal O}(\epsilon^2)\\
_2F_1\left(1+\epsilon,2; 2-\epsilon;\frac{p^2}{m^2}\right)&=& \frac{m^2}{m^2-p^2}\Bigg[
1-\epsilon \left( 1-\frac{p^2-m^2}{p^2}\ln\left(1-\frac{p^2}{m^2} \right) \right)
\Bigg]\nn\\
+{\cal O}(\epsilon^2)
\eea
\end{subequations}

\end{quote}

\section{Conclusions}~\label{conclu}

We have derived a non-perturbative expression for the scalar
propagator in SQED through its LKF transformation, starting from
its tree level expression in 3 and 4 space-time dimensions.
Equations~(\ref{LKFresults3D}) and~(\ref{LKFresult4D}) display two
of the main results of this paper. The LKFT of the scalar
propagator is written entirely in terms of basic functions of
momentum. Although our input is merely the bare propagator, its
LKFT, being non-perturbative in nature, contains useful
information of higher orders in perturbation theory. All the
coefficients of the $(\alpha \xi)^i$ at every order are correctly
reproduced without ever having to perform loop calculations. In
the weak coupling regime, LKFT results match onto the one-loop
perturbative results derived from the Lagrangian~(\ref{lag}) up to
gauge independent terms, a difference allowed by the structure of
the LKFT. This difference arises due to our approximate input, and
can be systematically removed at the cost of employing a more
complex input which would need to be calculated by the brute force
of perturbation theory.

\begin{acknowledgements}
The authors  wish to thank the SNI, CIC, and CONACyT grants. They
also acknowledge valuable discussions with A. Bashir.
\end{acknowledgements}

\section*{Appendix}

In this appendix we calculate the master integral~(\ref{master}).
Using the identity
\be
\frac{1}{A^{n}B^{p}}=\frac{\Gamma(n +
p)}{\Gamma(n)\Gamma(p)}\int_{0}^{1}dx\frac{x^{n - 1}(1 - x)^{p -
1}}{[xA + (1 - x)B]^{n + p}}\;,
\ee
we have
\be
J_{np}=\frac{\Gamma(n + p)}{\Gamma(n)\Gamma(p)}\int_{0}^{1}dx \int
d^{d}k\frac{x^{n - 1}(1 - x)^{p -1}}{[x(k - p)^{2} + (1 - x)(k^{2}
- m^{2})]^{n + p}} \;.\label{feynman}
\ee
We need to transform the denominator in a convenient form to use dimensional regularization results. Let
\bea
D&=&x(k - p)^{2} + (1 - x)(k^{2} - m^{2}) \nn \\
&=&x[k^{2} + p^{2} - 2k\cdot p] + (1 - x)k^{2} -m^{2}(1 - x) \nn \\
&=& k^{2} - 2k\cdot px + p^{2}x -m^{2}(1 - x) \;. \label{ec. 5.20}
\eea
 Now we perform the change of variables
 \be w=k - px\; \qquad
\Rightarrow \qquad k=w + px \;. \label{ec. 5.21} \ee Thus,
replacing Eq.~(\ref{ec. 5.21}) into Eq.~(\ref{ec. 5.20}) we have
that \bea
D&=&(w + px)^{2} - 2(w + px)\cdot px + p^{2}x -m^{2}(1 - x) \nn\\
&=&w^{2} + p^{2}x(1 - x) - m^{2}(1 - x)\;. \label{ec. 5.22}
\eea
Substituting Eq.~(\ref{ec. 5.22}), from Eq.~(\ref{feynman}) we have
\begin{eqnarray}
J_{np}=\frac{\Gamma(n+p)}{\Gamma(n)\Gamma(p)}\int
dx x^{n-1}(1-x)^{p-1}\int d^{d}w\frac{1}{[w^{2} + p^{2}x(1 - x) -
m^{2}(1 - x)]}\label{jnp}\;.
\end{eqnarray}
This expression is of the form \be \int\frac{d^{d}w}{(w^{2} +
s)^{n}} = i\pi^{d/2}\frac{\Gamma(n -
\frac{1}{2}d)}{\Gamma(n)}\frac{1}{s^{n - d/2}} \;,\label{ec.
A.1.8} \ee with $s=p^{2}x(1-x)-m^{2}(1-x)$, and therefore we can
write~(\ref{jnp}) as
 \bea J_{np}&=&\frac{(-1)^{n +
p}i\pi^{d/2}\Gamma(n + p -
\frac{d}{2})}{\Gamma(n)\Gamma(p)}\int_{0}^{1}dx [-p^{2}x(1 - x) +
m^{2}(1 - x)]^{\frac{d}{2} - n - p} \times \nn \\ && x^{n - 1}(1 -
x)^{p - 1} \;. \eea Performing the integration, \bea
\lefteqn{\int_{0}^{1}dx [-p^{2}x(1 - x) + m^{2}(1 - x)]^{\frac{d}{2} - n - p}x^{n - 1}(1 - x)^{p - 1}=} \nn \\
&=&\int_{0}^{1}x^{n - 1}(1 - x)^{p - 1}(1 - x)^{\frac{d}{2} - n - p}(-p^{2}x + m^{2})^{\frac{d}{2} - n - p}dx \nn \\
&=& (m^{2})^{\frac{d}{2} - n - p}\int_{0}^{1}x^{n - 1}(1 -
x)^{\frac{d}{2}- n - 1}(1 - \frac{p^{2}}{m^{2}}x)^{\frac{d}{2} - n
- p}dx
\eea
Next, we use
\be
\int_{0}^{1}x^{\lambda - 1}(1 - x)^{\mu - 1}(1 - \beta
x)^{-\nu}dx= B(\lambda,\mu){}_2F_{1}(\nu,\lambda;\lambda +
\mu;\beta)\;, \label{ec. 5.27}
\ee
where $\lambda=n$, $\mu=\frac{d}{2} - n$,
$\beta=\frac{p^{2}}{m^{2}}$ y $-\nu=\frac{d}{2} - n - p$. Thus,
\bea
\lefteqn{\int_{0}^{1}x^{n - 1}(1 - x)^{\frac{d}{2}- n - 1}(1 - \frac{p^{2}}{m^{2}}x)^{\frac{d}{2} - n - p}dx=} \nn \\
&=&B\bigg(n,\frac{d}{2} - n\bigg){}_2F_{1}\bigg(-\frac{d}{2} + n + p, n;\frac{d}{2};\frac{p^{2}}{m^{2}}\bigg) \nn \\
&=&\frac{\Gamma(n)\Gamma(\frac{d}{2} -
n)}{\Gamma(\frac{d}{2})}{}_2F_{1}\bigg(-\frac{d}{2} + n + p,
n;\frac{d}{2};\frac{p^{2}}{m^{2}}\bigg) \;.
\eea
Finally
\bea
J_{np}&=&(-1)^{n + p}i\pi^{d/2}(m^{2})^{\frac{d}{2} - n -
p}\frac{\Gamma(n + p - \frac{d}{2})\Gamma(\frac{d}{2} -
n)}{\Gamma(p)\Gamma(\frac{d}{2})}\times \nn \\ &&
{}_2F_{1}\bigg(-\frac{d}{2} + n + p,
n;\frac{d}{2};\frac{p^{2}}{m^{2}}\bigg) \;.
\eea

\end{document}